\documentclass[12pt,epsf]{article}

\usepackage[dvips]{graphicx}
\usepackage{latexsym}
\usepackage{epsfig}
\usepackage{amsfonts}

\begin{document}

\begin{center}
{\bfseries  \Large PIMC Simulations of Metal Hydrogen: Phase Transition and Equation of State} \vskip 5mm
 Alexander Novoselov$^\dag$
$^{*}$, Oleg Pavlovsky$^\dag$ $^\ddag$, Maxim Ulybyshev$^\dag$ $^\ddag$
\vskip 5mm {$\dag$ \small {\it Moscow State University, Moscow,
Russia} \vskip 1mm $\ddag$ \small {\it Institute for Theoretical
and Experimental Physics, Moscow, Russia} \vskip 1mm $*$ {\it
E-mail: novoselov@goa.bog.msu.ru} } \vskip 5mm
\end{center}

\begin{center}
\begin{minipage}{150mm}
\centerline{\bf Abstract}

The article is devoted to numerical studies of atomic (metal)
hydrogen with Path Integral Monte Carlo (PIMC) technique. The
research is focused on the range of temperatures and densities
where quantum statistics effects are crucial for electrons and
negligible for protons. In this range the equations of state are
obtained as a dependence of internal energy and pressure on
temperature and density. These dependences allow to detect and
describe the phase transition between solid and liquid phases.

\end{minipage}
\end{center}

\vskip 10mm

\section{Introduction}

One of the major recent achievements of astrophysics is the
discovery of numerous exoplanetary systems. Almost thousand such
planets have been discovered \cite{exopl_en}. Most of them are gas
giants up to ten Jovian masses. That is the reason that attracts
an increasing interest to the models of planetary evolution. By the
current conception gas giants mainly consist of hydrogen and
helium. So the equation of state of these elements is crucial for
the models of planetary formation and evolution.

The detection of huge magnetic moment of the solar system gas giants
has proven that they have liquid metal hydrogen core
\cite{exopl_mag}. There are some exoplanets that are much more
massive and maybe colder than Jupiter. Because of higher pressure
and less temperature their cores may contain not only liquid, but
also solid crystal hydrogen. The formation, evolution and
properties of planets are determined by the balance of gravity and
pressure, and the pressure in one's turn is determined by the
equation of state and thermodynamical parameters of the planetary
matter. In the cores of gas giants the prevailing substance is
metal hydrogen. It can be described as a many-body quantum system.
Its analytic analysis is extremely complicated, so the numerical
calculations are actual in this problem.

This article is devoted to Path Integral Monte Carlo simulation of
metal hydrogen. In the explored range of temperatures and
densities electrons form a degenerate quantum gas while nuclei can
be examined with classical statistics, that allows to avoid
fermion statistics problem. The parameters to be explored are
internal energy and pressure and their dependence on temperature
and density. We also focus on the phase transition between liquid
and crystal phases. It is detected and explored in a wide range of
densities.

It should be noted that the study of metal hydrogen is important
not only for astrophysics, but also due to the progress in diamond
anvil cell experiments that have recently obtained crystal metal
hydrogen in the laboratory \cite{nmat}.

We broadly use nuclear units in this work: $k_e=\hbar=e=m_p=1;$
here $k_e$ is Coulomb constant and $m_p$ is proton mass.
Corresponding units of principal physical quantities are: nuclear
Bohr radius $a_{0N}=L_N=2.9\times10^{-14}\ \mathrm{m}$ for length,
nuclear Hartree $Ha=E_N=8.0\times10^{-15}\ \mathrm{J}$ for energy,
$p_N=3.3\times10^{26}\ \mathrm{Pa}$ for pressure unit,
$\rho_N=7.0\times10^{13}\ \mathrm{kg/m^3}$ for density unit and
$T_N=5.8\times10^8\ \mathrm{K}$ for temperature unit. In nuclear
units electron mass is $m_e=5.4\times10^{-4}$ and (electron) Bohr
radius is $a_{0e}=1/m_e=1.8\times10^3.$

We study the dependence of the atomic hydrogen properties on temperature and density, described by parameters $\beta$
\begin{equation}
\beta=1/k_B T
\label{beta}
\end{equation}
and $r_s$ (Wigner-Seitz radius)
\begin{equation}
\rho={m\over {4\over3}\pi r_s^3}
\label{r_s}
\end{equation}
respectively. We simulate a finite cell of the substance,
containing $N_p=128$ particles. The properties to be evaluated in
the simulation are internal energy (the sum of kinetic and
potential energies of the particles)
\begin{equation}
E=K+V
\label{ETV}
\end{equation}
and pressure $P.$ It is well known that the functions $E(\rho,T)$ and
$P(\rho,T)$ provide a complete thermodynamical description of the
system. In order to obtain an obvious measure of the order of the
system we also calculate Lindemann ratio
\begin{equation}
\mathcal{L}={\sqrt{\langle x^2\rangle}\over R_n}. \label{Lind_R}
\end{equation}
Here $\langle x^2\rangle$ is the particle displacement from its
site in crystal lattice and $R_n$ is the distance to the nearest neighbouring
particle. Lindemann ratio is used to explicitly distinguish
chaotic and crystal phase.

\section{Model}


The Hamiltonian of atomic hydrogen is
\begin{equation}
H_{full}=K_N+K_e+V_0+V_e+V_{int}.
\label{full_H}
\end{equation}
Here $K_N$ and $K_e$ are kinetic energies of nuclei (protons) and
electrons respectively. $V_0,$ $V_e$ and $V_{int}$ are potential
energies of nuclei-nuclei, electron-electron and nuclei-electron
interaction respectively; all three are sums of pair Coulomb
interaction, for example
\begin{equation}
V_0=\sum_{i_1=1}^{N_p}\sum_{i_2=1}^{i_1-1}{1\over r_{i_1i_2}}.
\label{V_0}
\end{equation}

There is a wide range of temperatures and densities where on the
one hand electrons can be considered as degenerate Fermi gas and
Tomas-Fermi model is applicable to them (i. e. Fermi statistics is
of primary importance), but on the other hand protons are strongly
not degenerate and their statistics is of no importance. On these
assumptions we can deal only with protons, moreover we can use
classical (Botzmann) statistics. The effect of taking electrons
into account is Thomas-Fermi screening. So the effective
Hamiltonian is
\begin{equation}
H_{full}=K_N+V_N.
\label{H}
\end{equation}
Here $V_N$ is potential energy of protons with screened interaction:
\begin{equation}
V_N=\sum_{i_1=1}^{N_p}\sum_{i_2=1}^{i_1-1}{\exp\{-{r_{i_1i_2}/R_{TF}}\}\over r_{i_1i_2}}.
\label{V_N}
\end{equation}
Thomas-Fermi screening length $R_{TF}$ is given by
\begin{equation}
R_{TF}=\root{3}\of{\pi\over12}\sqrt{a_{0e}r_s}.
\label{R_TF}
\end{equation}


The Hamiltonian (\ref{full_H}) can be reduced to (\ref{H}) under
following conditions. First, we want to neglect effects of nuclear forces for protons, so their separation (which is
approximately $r_s$) must me much greater than their size $R_p$.
Second, we want to applicate Thomas-Fermi theory to electrons,
that can be done if there are many electrons within screening
length. This leads intuitively obvious restriction that nuclei
separation must me less than (electron) Bohr radius. So, our
approximation are applicable for densities corresponding
\begin{equation}
R_p\ll r_s\ll a_{0e}.
\label{DoA_Rs}
\end{equation}
The limits in nuclear and SI units are $R_p\approx
3\times10^{-2}\approx 9\times10^{-16}\ \mathrm{m}$ and
$a_{0e}\approx 2\times10^3\approx 5\times10^{-11}\ \mathrm{m}.$
The estimations for limiting densities (\ref{r_s}) are
$\rho_{min}\approx 3\times10^3\ \mathrm{kg/m^3}$ and
$\rho_{max}\approx 6\times10^{17}\ \mathrm{kg/m^3}.$

Next, our approximation is valid if electrons are degenerate and
protons are not. The degeneracy temperature can be estimated as
$\beta_d\approx m r_s^2.$ So, the acceptable range of temperatures
depends on density and it is defined as
\begin{equation}
m_e r_s^2\ll\beta\ll r_s^2.
\label{DoA_beta}
\end{equation}
The temperature limits in nuclear units are $\beta_{min}\approx
5\times10^{-4}\ r_s^2$ and $\beta_{max}\approx r_s^2.$ This leads
following estimations at given densities (in SI units):
$T_{min}\approx 0.9\rho^{2\over3}\mathrm{kg^{-2/3}m^2K}$ and
$T_{max}\approx 2\times10^5\rho^{2\over3}\mathrm{kg^{-2/3}m^2K}.$

\section{PIMC}

\subsection{Path Integral Monte Carlo}

Suppose a system, determined by coordinates $\mathbf{x},$ in
imaginary time. The density matrix of such system with Hamiltonian
$H$ at the temperature $\beta$ is
\begin{equation}
\rho_{x_0\to x_{N_t}}=\langle\mathbf{x}_0|e^{-\beta H}|\mathbf{x}_{N_t}\rangle.
\label{rho}
\end{equation}
Its partition function is
\begin{equation}
Z=\mathrm{tr}\rho=\int d\mathbf{x_0}\langle\mathbf{x}_0|e^{-\beta H}|\mathbf{x}_0\rangle.
\label{Z_0}
\end{equation}
Average observable $A$ is calculated with
\begin{equation}
\langle A\rangle={1\over Z}\mathrm{tr}(A\rho)={1\over Z}\int d\mathbf{x}_0\langle\mathbf{x}_0|Ae^{-\beta H}|\mathbf{x}_0\rangle.
\label{A_0}
\end{equation}

To proceed to the path integral formulation, introduce the "time step" $\tau,$ defined as
\begin{equation}
1/T=\beta=N_t\tau.
\label{Na}
\end{equation}
and decompose the density matrix into a product of $N_t$ density matrices
\begin{equation}
\rho_{x_0\to x_{N_t}}=\rho_{x_0\to x_{1}}\ldots\rho_{x_{t-1}\to x_{t}}\ldots\rho_{x_{N_t-1}\to x_{N_t}},
\label{rho_decomp}
\end{equation}
where each of these intermediate matrices is
\begin{equation}
\rho_{x_{t-1}\to x_{t}}=\langle \mathbf{x}_{t-1}|e^{-\tau H}|\mathbf{x}_{t}\rangle\equiv e^{-S_t}.
\label{rho_decomp}
\end{equation}
"Lattice action" $S$ is defined as
\begin{equation}
S=\sum_{t=1}^{N_t}S_t.
\label{rho_decomp}
\end{equation}
In fact the above decomposition given by Trotter formula is
correct only if $N_t\to\infty,$ and for real simulation $N_t$ will
be chosen large enough to eliminate the dependence of the result
on it. Next we introduce the notation
\begin{equation}
\mathcal{D}\mathbf{x}=\prod_{t=1}^{N_t}d\mathbf{x}_{t}
\label{Dx}
\end{equation}
and consequently the formulae (\ref{Z_0}) and (\ref{A_0}) can be
represented as follows:
\begin{equation}
Z=\int\mathcal{D}\mathbf{x}e^{-S},
\label{Z}
\end{equation}
\begin{equation}
\langle A\rangle={\int\mathcal{D}\mathbf{x}Ae^{-S}\over\int\mathcal{D}\mathbf{x}e^{-S}}=\int A{\mathcal{D}\mathbf{x}e^{-S}\over\int\mathcal{D}\mathbf{x}e^{-S}}
\label{A}
\end{equation}

Formula (\ref{A}) reveals the idea of Path Integral Monte Carlo.
Since we have a (large enough) set of paths
$x={\mathbf{x}_{0}\ldots\mathbf{x}_{t}\ldots\mathbf{x}_{N_t}},$
where the probability of the path to be included into the set is
proportional to its "statistical weight"
\begin{equation}
\pi(x)\sim e^{-S(x)}.
\label{P_pi}
\end{equation}
The average of any observable can be measured by simple
(arithmetic) averaging over this set.

\subsection{Algorithms}

The way to obtain properly distributed (\ref{P_pi}) paths is based
on the property of Markov chains to converge to the limiting
distribution. A sufficient condition of the convergence to the
limiting distribution $\pi(x)$ for the Markov chain with a
transition probability $\mathcal{P}(x\to x')$ is the detailed
balance condition:
\begin{equation}
\mathcal{P}(x\to x')\pi(x)=\mathcal{P}(x'\to x)\pi(x').
\label{det bal}
\end{equation}
The specific form of $\mathcal{P}(x\to x')$ is not fixed, but it
must be constructed carefully as it crucially affects the time of
"thermalization" (convergence to the limiting distribution).

A generalized Metropolis-Hastings algorithm is based on
the decomposition of transition probability:
\begin{equation}
\mathcal{P}(x\to x')=\mathcal{T}(x\to x')\mathcal{A}(x\to x')+\delta(x-x')\left\{1-\int dy\mathcal{T}(x\to
y)\mathcal{A}(x\to y)\right\},\label{metro_gen}
\end{equation}
here
\begin{equation}
\mathcal{A}(x\to x')=\min\left[1,{\mathcal{T}(x'\to x)\pi(x')\over \mathcal{T}(x\to
x')\pi(x)}\right].\label{metro_A_gen}
\end{equation}
It satisfies the detailed balance condition for any
$\mathcal{T}(x\to x').$ Formula (\ref{metro_A_gen}) means the
following. First, generate a new (trial) configuration with
probability $\mathcal{T};$ then accept it (add it to the set) with
probability $\mathcal{A}$ or reject it (return to the previous
configuration and add an other copy of it to the set) with
probability $1-\mathcal{A}.$ The specific form of the algorithm is
defined by the choice of the function $\mathcal{T}(x\to x').$ The
theoretically best choice is "heat bath":
\begin{equation}
\mathcal{T}(x\to x')=\pi(x'),\
\mathcal{A}(x\to x')=1,\
\mathcal{P}(x\to x')=\pi(x').
\label{heat_bath}
\end{equation}
Unfortunately, most probability distributions can not be generated
directly fast enough, so we have to use a general type of the
algorithm (\ref{metro_A_gen}). There are two demands to the
distribution $\mathcal{T}(x\to x'):$ first, it must be close to
$\pi(x'),$ second, there must be an algorithm of generating it
numerically very fast. It is rather natural to choose $\mathcal{T}(x\to x')$ as the
kinetic part of the "statistical weight", then the acceptance
probability $\mathcal{A}(x\to x')$ is proportional to its
potential part.

Primitive algorithm is based on "sweep" when the transition from
"old" configuration to "new" one is a try to change only one
coordinate (or coordinates in the only imaginary time slice $t$).
For large systems and for large number of slices it has huge
autocorrelation. It means that "new" configurations turn out to
look like "old", and it takes much time to obtain really
statistically independent ones. This problem can be solved with the
multilevel algorithm \cite{Ceperley}. It is based on fast
generation of a rough approximation of the path, that increase the
acceptance rate of the further more accurate one.

Consider a bisection multilevel algorithm. We start from a part of the path with length $2^{N_{level}}$ slices, for example $s=({\mathbf x_0},\ldots,{\mathbf x_{2^{N_{level}}}}).$ This part of the path is divided into levels $s_k.$ Zero level consists of the coordinates on the boundaries of the chosen part of the pass: $s_0=({\mathbf x_0},{\mathbf x_{2^{N_{level}}}})$. They are not to be changed during the current multilevel update. The first level consist of the coordinates on one middle time slice  $s_1=({\mathbf x_{2^{N_{level}-1}}}).$ The second level consist of two time slices $s_2=({\mathbf x_{2^{N_{level}-2}}},{\mathbf
x_{2^{N_{level}-1}+2^{N_{level}-2}}}),$ etc. There are $2^{k-1}$
slices in the $k$-th level. Introduce "level action"
$\pi_k(s_k)\equiv\pi_k(s_0,\ldots,s_{k-1},s_k),$ which is a
function of $s_k$ and previous levels coordinates are parameters.
Intermediate levels actions can be chosen arbitrary, the only
requirement is that the action of the last level must be the
lattice action:
\begin{equation}
\pi_{N_{level}}(s_{N_{level}})=\pi(s).
\label{pi_level_cond}
\end{equation}
Then start a kind of Metropolis-Hastings algorithm with trial probability distribution
$$
\mathcal{T}_k(s_k')=\mathcal{T}_k(s_0',\ldots,s_{k-1}';s_k;s_{k+1},\ldots,s_{N_{level}}\to
s_0',\ldots,s_{k-1}';s_k';s_{k+1},\ldots,s_{N_{level}})
$$
and acceptance probability
\begin{eqnarray}
&\mathcal{A}_k(s_k')=\mathcal{A}(s_0',\ldots,s_{k-1}';s_k;s_{k+1},\ldots,s_{N_{level}}\to
s_0',\ldots,s_{k-1}';s_k';s_{k+1},\ldots,s_{N_{level}})=\nonumber\\
&=\min\left[1,{\mathcal{T}_k(s_k)\pi_k(s_k')\pi_{k-1}(s_k)\over
\mathcal{T}_k(s_k')\pi_k(s_k)\pi_{k-1}(s_k')}\right].
\label{ml_A}
\end{eqnarray}
It satisfies the level detailed balance condition
\begin{equation}
\mathcal{P}_k(s_k'){\pi_k(s_k)\over\pi_{k-1}(s_{k-1})}=\mathcal{P}_k(s_k){\pi_k(s_k')\over\pi_{k-1}(s_{k-1}')}
\label{ml_det_bal}
\end{equation}
that leads to full detailed balance (\ref{det bal}):
\begin{equation}
\pi_k(s_k)=\int ds_{k+1}\ldots
ds_{N_{level}}\pi(s).\label{pi_level_int}
\end{equation}

\subsection{Some Details}

Our simulation is limited in the number of particles, and
consequently in the spatial size of the cell. We use cubic cell
and periodic boundary conditions in space. The size of the cell is
\begin{equation}
L=\root{3}\of{{4\over3}\pi N}\ r_s.
\label{L}
\end{equation}

The $\alpha=x,y,z$ coordinate of $i$-th particle in the
$t$-th time slice is denoted by $x_i^\alpha(t)$. To describe the
configuration completely we also need "winding numbers"
$n_i^\alpha(t)=-1,0,1,$ that denote if the corresponding path
"skips" from one side of the cell to another through periodic
spatial boundary conditions. Potential energy of particle
interaction (particles can be in different "copies" of the cell
due to boundary conditions) is determined by their separation
\begin{equation}
r_{i_1i_2}^{n^1n^2n^3}(t)=\sqrt{ \sum_{\alpha=1}^{3} (x_{i1}^\alpha(t)-x_{i2}^\alpha(t)+Ln^\alpha)^2 }.
\label{r_separation}
\end{equation}
In the notation, described above, the lattice action corresponding to
the Hamiltonian (\ref{H}) with potential energy (\ref{V_N}) and
periodic spatial boundary conditions is set as
\begin{equation}
-\mathrm{ln}\pi=S=S_T+S_V.
\label{S}
\end{equation}
\begin{equation}
S_T=\sum_{t=1}^{N_t}\sum_{i=1}^{N_p}\sum_{\alpha=1}^{3}{(x_i^\alpha(t)-x_i^\alpha(t-1)+Ln_i^\alpha(t))^2\over2\tau},
\label{S_T}
\end{equation}
\begin{equation}
S_V=\sum_{t=1}^{N_t}\sum_{i_1=1}^{N_p}\sum_{i_2=1}^{i_1}\sum_{n_{i_1i_2}^{1,2,3}(t)=-1}^{1} {\exp\{-{r_{i_1i_2}^{n^1n^2n^3}(t)/R_{TF}}\}\over r_{i_1i_2}^{n^1n^2n^3}(t)} \tau.
\label{S_V}
\end{equation}

In the case of periodic boundary conditions the trial probability
density based on the kinetic part of the action can be represented as
(skipping irrelevant indices for simplicity)
\begin{equation}
\mathcal{T}(x^\infty(t)|n(t+1)-n(t))\sim\exp\left\{-{1\over
\tau}\left[x^\infty(t)-{x(t+1)+x(t-1)+L(n(t+1)-n(t))\over2}\right]^2\right\}.
\label{T_p_x}
\end{equation}
Gaussian (it has infinite range) distribution of
$x^\infty(t)\equiv x'(t)+Ln'(t)$ can be generated fast (we use
Box-Muller transform) and allows to determine $n'(t),$ $n'(t+1)$
and $x'(t)$ due to conditions $-L/2<x'(t)<L/2$ and
$n'(t+1)-n'(t)=n(t+1)-n(t).$

We use multilevel algorithm. Though the level action can be chosen
arbitrary, there is a theoretically optimal choice. The action of
the level should be obtained by integrating out the next levels
coordinates in the full lattice action:
\begin{equation}
\pi_k(s_k)=\int ds_{k+1}\ldots ds_{N_{level}}\pi(s).
\label{pi_level_int}
\end{equation}
For our model with action (\ref{S}),(\ref{S_T}),(\ref{S_V}) it
leads to a quite simple and effective algorithm. Trial probability
distribution for each bisection is (\ref{T_p_x}), where the level
time step is $\tau\to\tau_k=2^{N_{level}-k}\tau$ and winding
number conserves $n_k(t+1)-n_k(t)=n_{k-1}(t).$ This trial
distribution together with the condition (\ref{pi_level_int})
leads to the acceptance probability
\begin{equation}
\mathcal{A}_k(s_k')=\min\left[1,{e^{-S_V(s_k')}\over
e^{-S_V(s_k)}}\right],
\label{ml_A_V}
\end{equation}
$S_V(s_k)$ is determined by (\ref{S_V}) with the first sum only
over the slices that belong to the level $s_k.$ $\tau$ is not level time step (as it was in the kinetic part) but the real
time step.

\section{Results}

The calculations were performed for following parameters.
$Na=1/\beta$ from $0.5\times10^{-5}$ to $4.75\times10^{-5}$ with step
$0.25\times10^{-5}$ and for additional points $0.57\times10^{-5}$ and
$0.66\times10^-5.$ $r_s$ was changed from $200$ to $450$ with step $50.$ These
values in nuclear units correspond to SI values of temperature
from $2.9\times10^3\ \mathrm{K}$ to $27.7\times10^3\ \mathrm{K}$
and density from $183\times10^3\ \mathrm{kg/m^3}$ to
$2085\times10^3\ \mathrm{kg/m^3}.$ The lattice of calculation
points will be shown in Figure \ref{phase_plane} (discussed
later).

\subsection{Energy}

Average internal energy $\langle E\rangle$ is calculated as
(\ref{ETV}), taking into account
(\ref{Na}),(\ref{S_T}),(\ref{S_V}):
\begin{equation}
\langle V\rangle=\langle{S_V\over\beta}\rangle,
\label{obs_V}
\end{equation}
\begin{equation}
\langle K\rangle=\langle{3N_p\over2\tau}-{S_T\over\beta}\rangle.
\label{obs_T}
\end{equation}
Note that while the potential energy observable is rather
intuitive, the kinetic energy one is quite different from
intuitive (but incorrect) form. By the way in real numerical calculations the averaging should be done exactly as in (\ref{obs_T}). $\langle T\rangle=3N_p/2\tau-\langle S_T\rangle/\beta$ seems similar but leads to large errors because of substraction of very close large numbers.

\begin{figure}[h!]
\begin{center}
\epsfysize=100mm\epsfbox{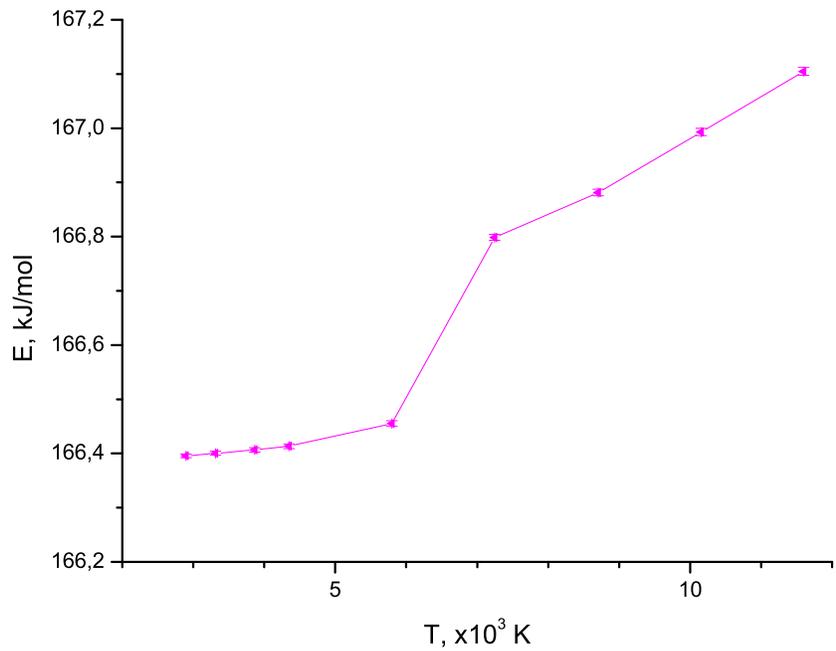}
\end{center}
\caption{$E(T)$ at $\rho=183\times10^3\ \mathrm{kg/m^3}\ (r_s=450)$}
\label{E_450}
\end{figure}

\begin{figure}[h!]
\begin{center}
\epsfysize=100mm\epsfbox{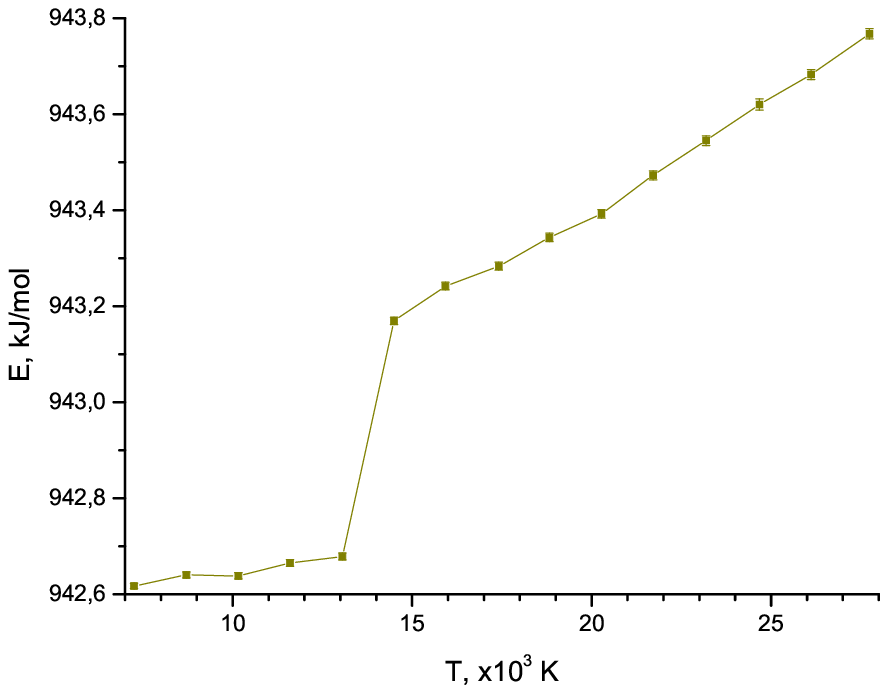}
\end{center}
\caption{$E(T)$ at $\rho=2085\times10^3\ \mathrm{kg/m^3}\ (r_s=200)$}
\label{E_200}
\end{figure}

\begin{figure}[h!]
\begin{center}
\epsfysize=100mm\epsfbox{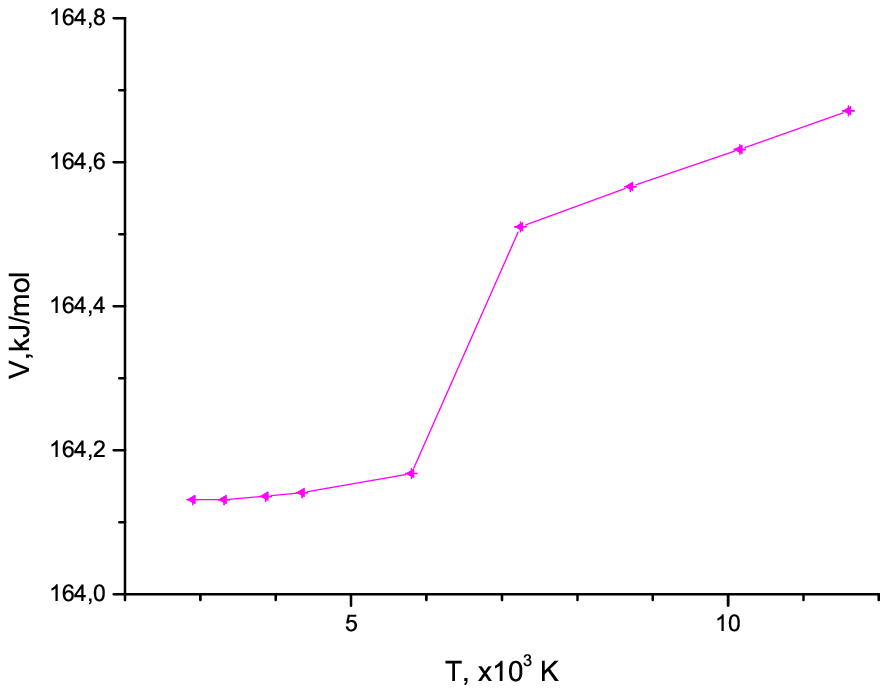}
\end{center}
\caption{$V(T)$ at $\rho=183\times10^3\ \mathrm{kg/m^3}\ (r_s=450)$}
\label{V_450}
\end{figure}

\begin{figure}[h!]
\begin{center}
\epsfysize=100mm\epsfbox{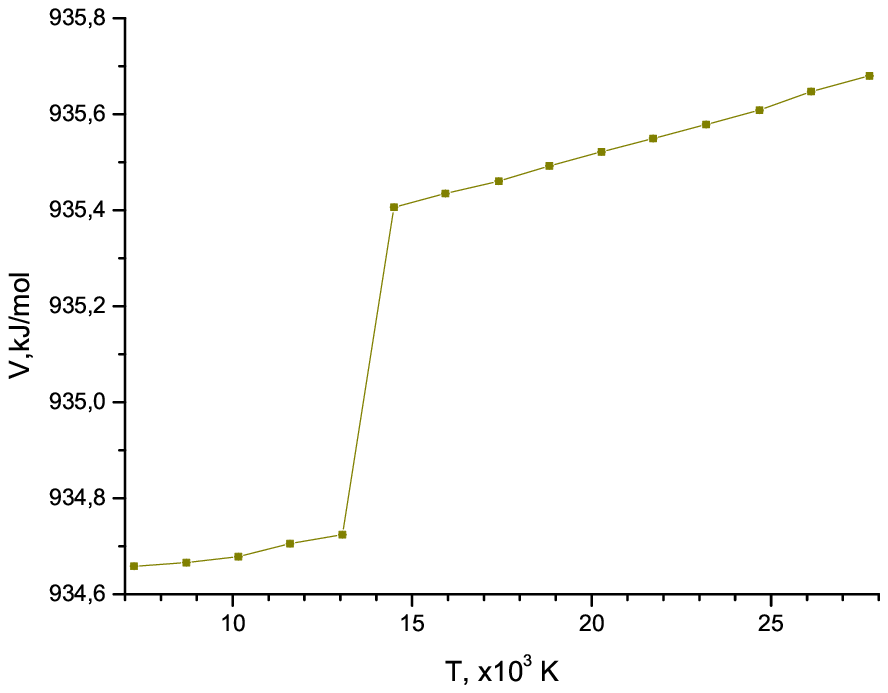}
\end{center}
\caption{$V(T)$ at $\rho=2085\times10^3\ \mathrm{kg/m^3}\ (r_s=200)$}
\label{V_200}
\end{figure}

\begin{figure}[h!]
\begin{center}
\epsfysize=100mm\epsfbox{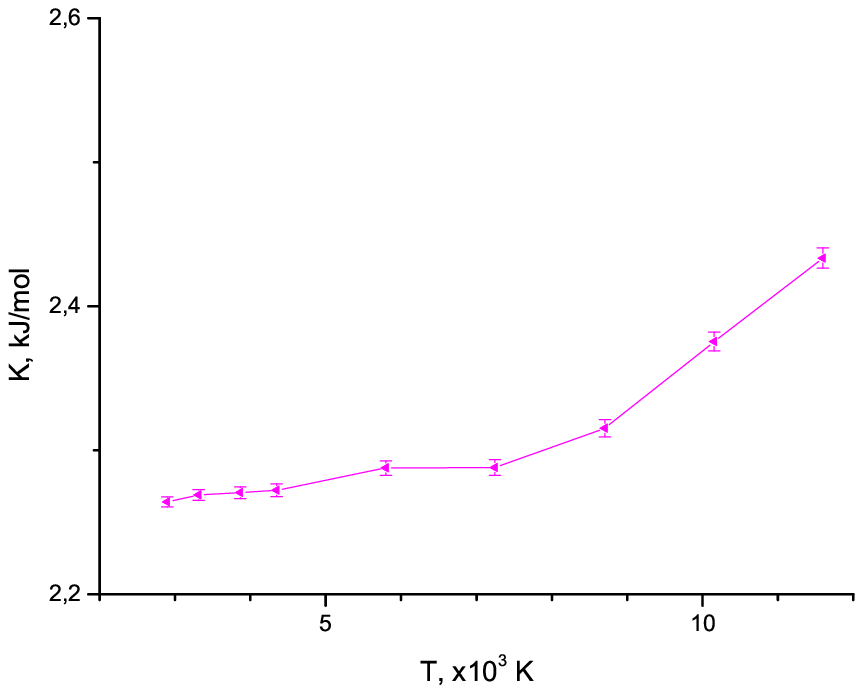}
\end{center}
\caption{$K(T)$ at $\rho=183\times10^3\ \mathrm{kg/m^3}\ (r_s=450)$}
\label{T_450}
\end{figure}

\begin{figure}[h!]
\begin{center}
\epsfysize=100mm\epsfbox{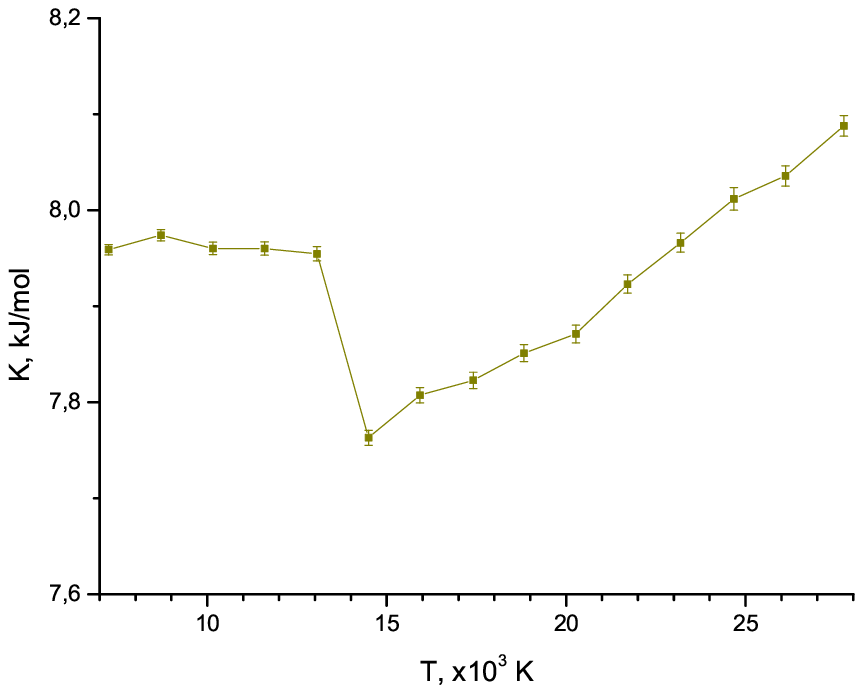}
\end{center}
\caption{$K(T)$ at $\rho=2085\times10^3\ \mathrm{kg/m^3}\ (r_s=200)$}
\label{T_200}
\end{figure}

Figures \ref{E_450} and \ref{E_200} show the internal energy $E$
as a function of temperature for the densities $183\times10^3\
\mathrm{kg/m^3}$ and $2085\times10^3\ \mathrm{kg/m^3}$
respectively. In both cases we observe a slight increase with
increasing temperature and an acute jump at certain temperature
that is associated with the phase transition. Figures \ref{V_450}
and \ref{V_200} show the potential energy at these densities,
which behaves similar to full energy, i. e. increases and has a
jump up at the same temperatures for each given density. Figures
\ref{T_450} and \ref{T_200} show the kinetic energy at these
densities. Its behaviour is different from potential and full
energy. It also increases, but jumps {\it down} at phase
transition. At lower densities this jump vanishes and turns into a
jump of the slope only. So, it looks like a second-order phase
transition at densities $261\times10^3\ \mathrm{kg/m^3}$ and
lower and like a first order phase transition at densities
$618\times10^3\ \mathrm{kg/m^3}$ and higher.

\begin{figure}[h!]
\begin{center}
\epsfysize=100mm\epsfbox{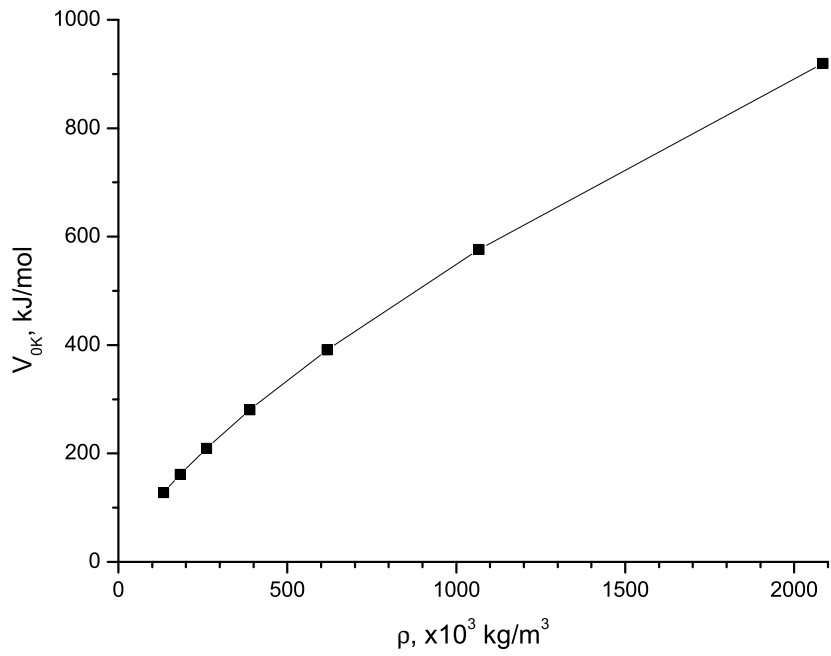}
\end{center}
\caption{$V_{0K}(\rho)$}
\label{V_0K}
\end{figure}

\begin{figure}[h!]
\begin{center}
\epsfysize=100mm\epsfbox{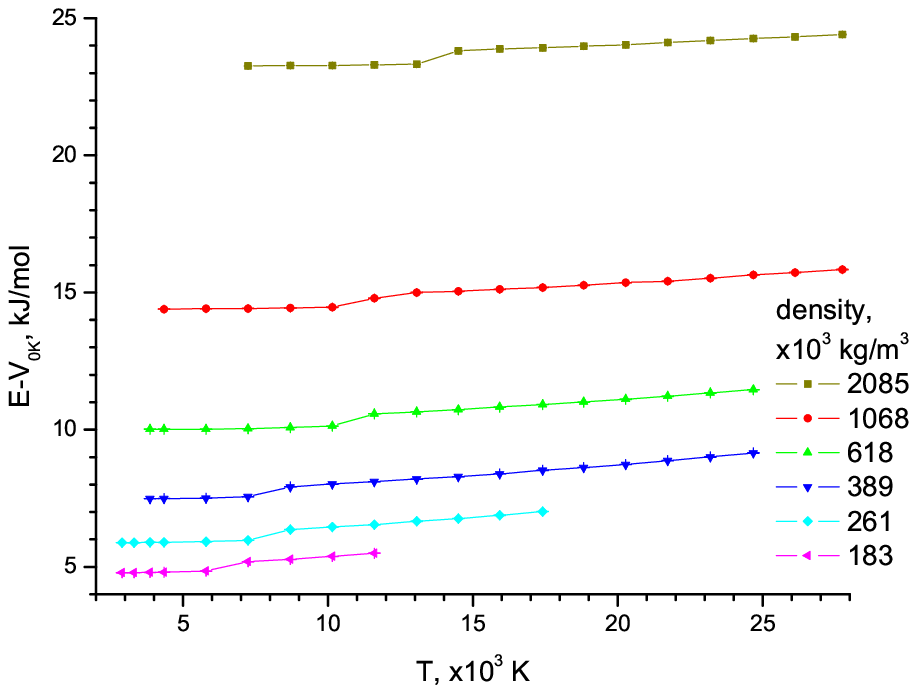}
\end{center}
\caption{$E(T)-V_{0K}$ at different densities}
\label{E_all}
\end{figure}

\begin{figure}[h!]
\begin{center}
\epsfysize=100mm\epsfbox{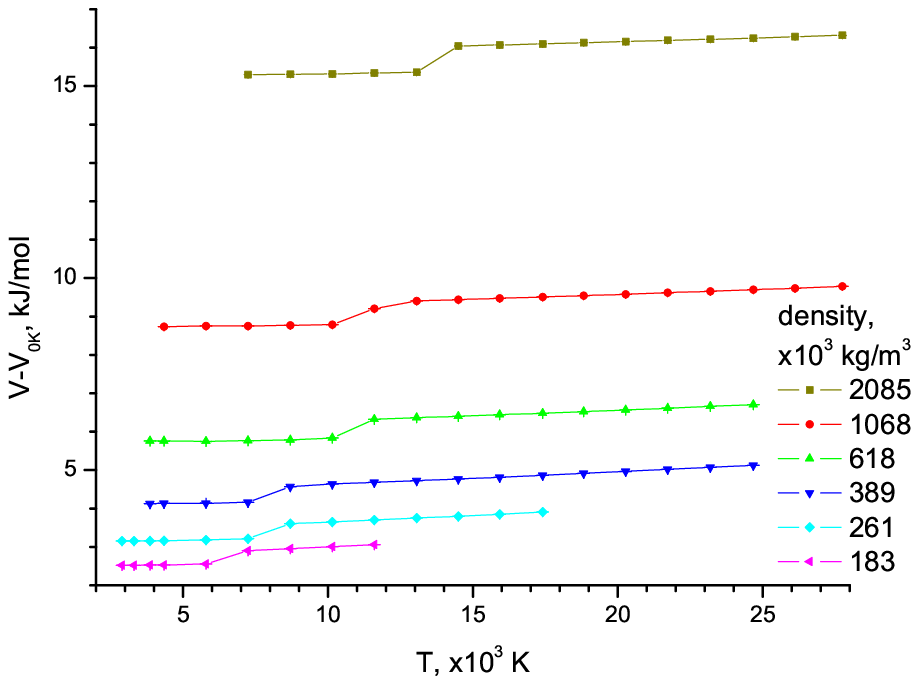}
\end{center}
\caption{$V(T)-V_{0K}$ at different densities}
\label{V_all}
\end{figure}

\begin{figure}[h!]
\begin{center}
\epsfysize=100mm\epsfbox{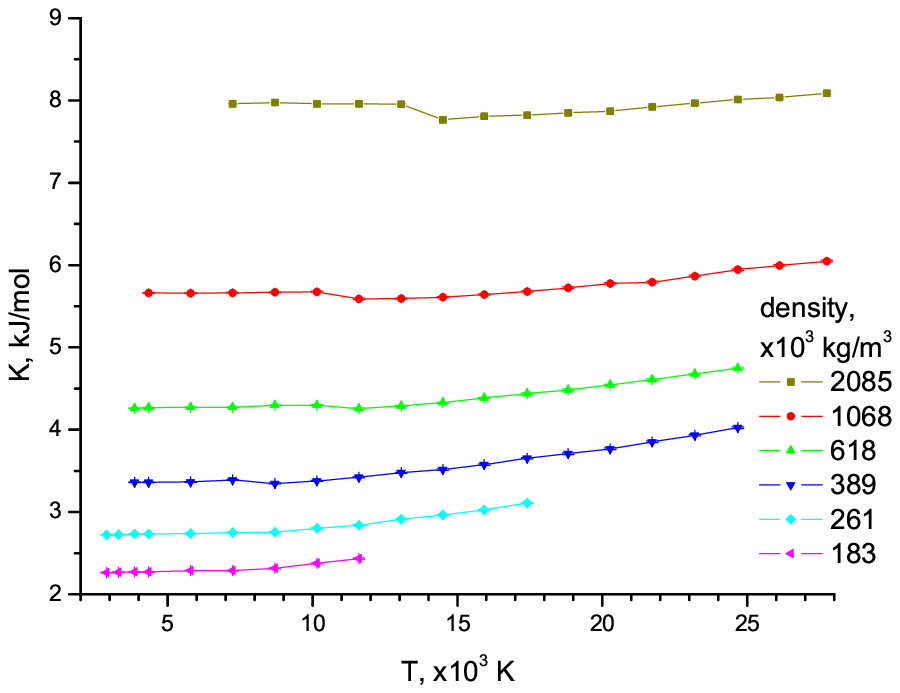}
\end{center}
\caption{$K(T)$ at different densities}
\label{T_all}
\end{figure}

We can see that the properties of
the system depend on density much stronger than on temperature.
Correspondingly, the internal energy almost totally consist of
potential energy determined by the distance between protons i. e.
by density. In spite of this fact, the jumps of both parts of
energy at the phase transition are of close magnitudes. In order to extract the main term we introduce $V_{0K}$ - potential energy of "ideal zero temperature" crystal. It means that the particles in this crystal are exactly in the sites of its bcc (body-centric cubic) lattice (in all time slices).
$V_{0K}$ depends only on density and this dependence is shown in the Figure \ref{V_0K}.
We substract this zero energy from full and potential energy in order to extract non-trivial terms. It turns out that substracted full and potential energy and kinetic energy are of the same magnitude; their dependences on density and temperature also have close magnitudes.
The substracted full internal, substracted potential and kinetic energies for a range of densities between
$183\times10^3\ \mathrm{kg/m^3}$ and $2085\times10^3\
\mathrm{kg/m^3}$ are shown in figures \ref{E_all}, \ref{V_all} and
\ref{T_all} respectively.

\subsection{Pressure. Equation of state}

The observable for pressure is
\begin{equation}
\langle P\rangle={2\over3L^3}\left(\langle K\rangle-{1\over2}\langle\sum_{i<j}{\partial
V\over\partial{\mathbf r}_{ij}}{\mathbf r}_{ij}\rangle\right).
\label{P}
\end{equation}

\begin{figure}[h!]
\begin{center}
\epsfysize=100mm\epsfbox{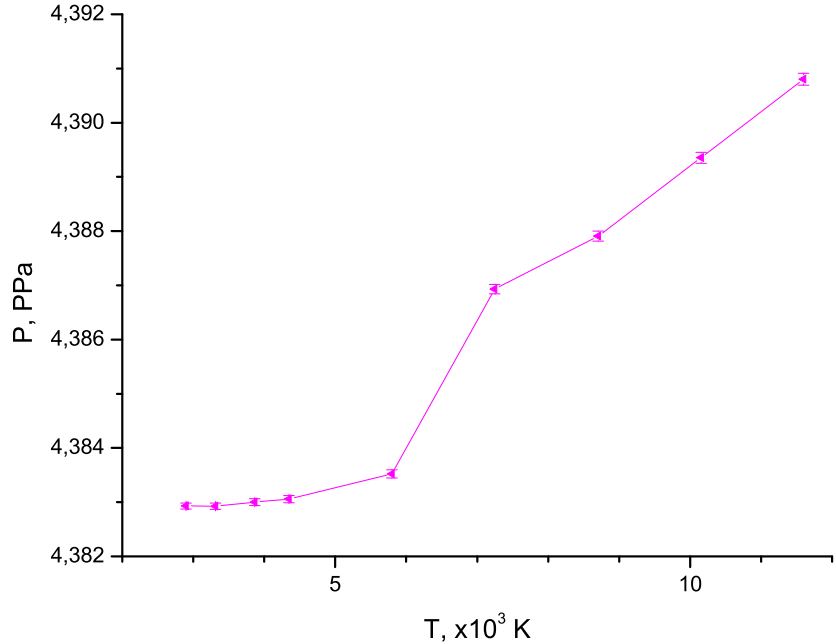}
\end{center}
\caption{$P(T)$ at $\rho=183\times10^3\ \mathrm{kg/m^3}\ (r_s=450)$}
\label{P_450}
\end{figure}

\begin{figure}[h!]
\begin{center}
\epsfysize=100mm\epsfbox{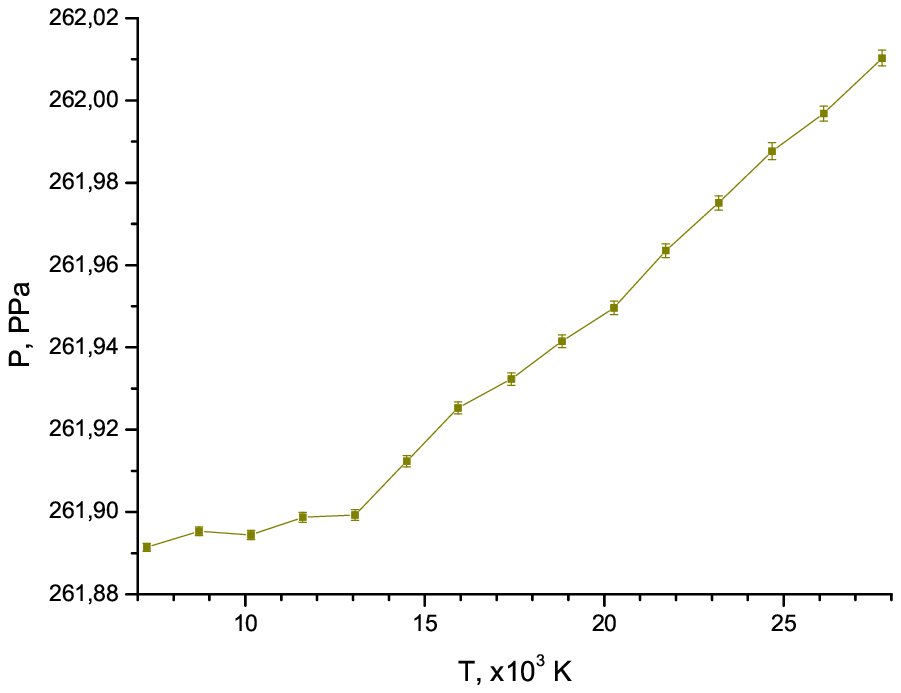}
\end{center}
\caption{$P(T)$ at $\rho=2085\times10^3\ \mathrm{kg/m^3}\ (r_s=200)$}
\label{P_200}
\end{figure}

\begin{figure}[h!]
\begin{center}
\epsfysize=100mm\epsfbox{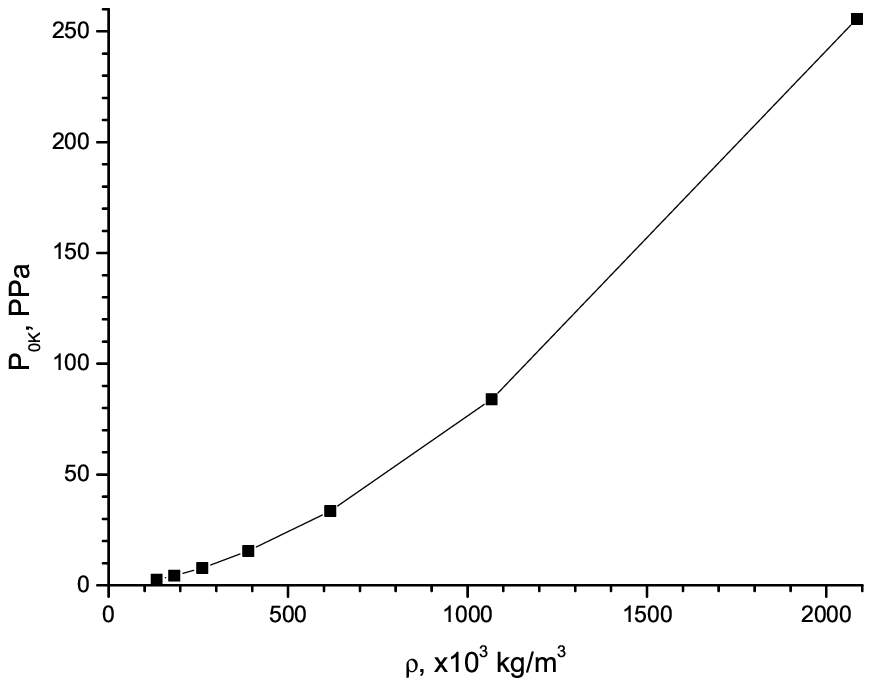}
\end{center}
\caption{$P(T)$ at different densities}
\label{P_0K}
\end{figure}

\begin{figure}[h!]

\begin{center}
\epsfysize=100mm\epsfbox{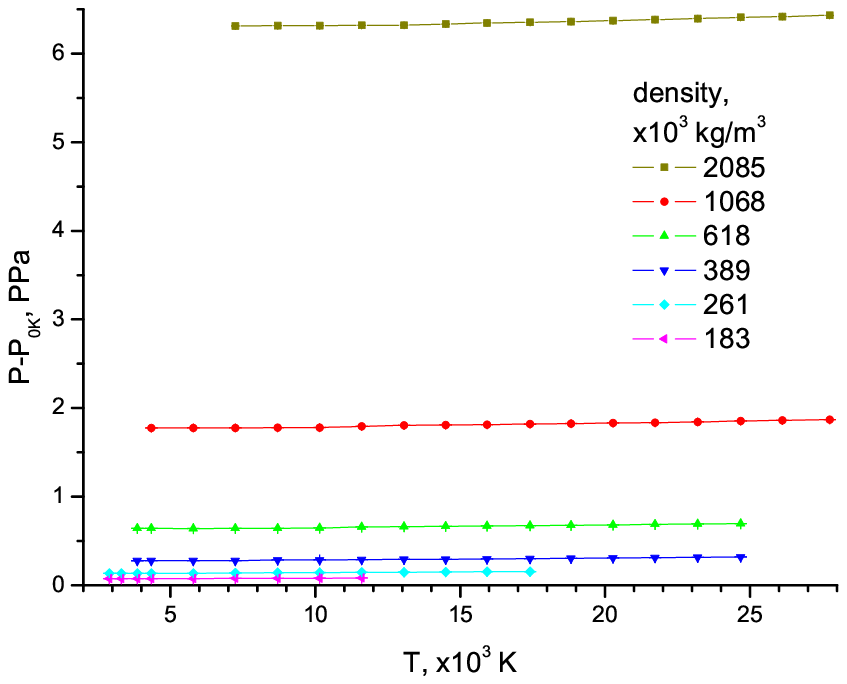}
\end{center}
\caption{$P(T)-P_{0K}$ at different densities} \label{P_all}
\end{figure}

Figures \ref{P_450} and \ref{P_200} show the temperature
dependence of pressure at above mentioned densities. It has a jump
at the same temperatures as energy that proves the existence of
phase transition. The expression (\ref{P}) allows to determine $P_{0K}$ similar to $V_{0K}$ and perform similar substraction procedure. Figure (\ref{P_0K}) shows the dependence of $P_{0K}$ on density and
Figure \ref{P_all} shows the substracted pressure for all
the range of explored densities. Similar to energy, pressure mainly
depends on density and quite slightly changes with temperature. In
fact it is just what should be expected for condensed matter. The
disadvantage of this property is a trouble with thermodynamical
calculations due to orders of magnitude difference between partial
derivatives by density and by temperature. An intuitive
illustration can be seen in Figure \ref{P_all}. Formally the
function $P(\rho,T)$ allows to determine isobars, but the
resolution of experimental data is insufficient despite of quite
large number of points. We can only say that isobars are some
lines close to lines of constant density, but having some little
unknown slope. This problems can be solved in different ways, but
they are not to be discussed here.


\subsection{Phase transition}

\begin{figure}[h!]
\begin{center}
\epsfysize=100mm\epsfbox{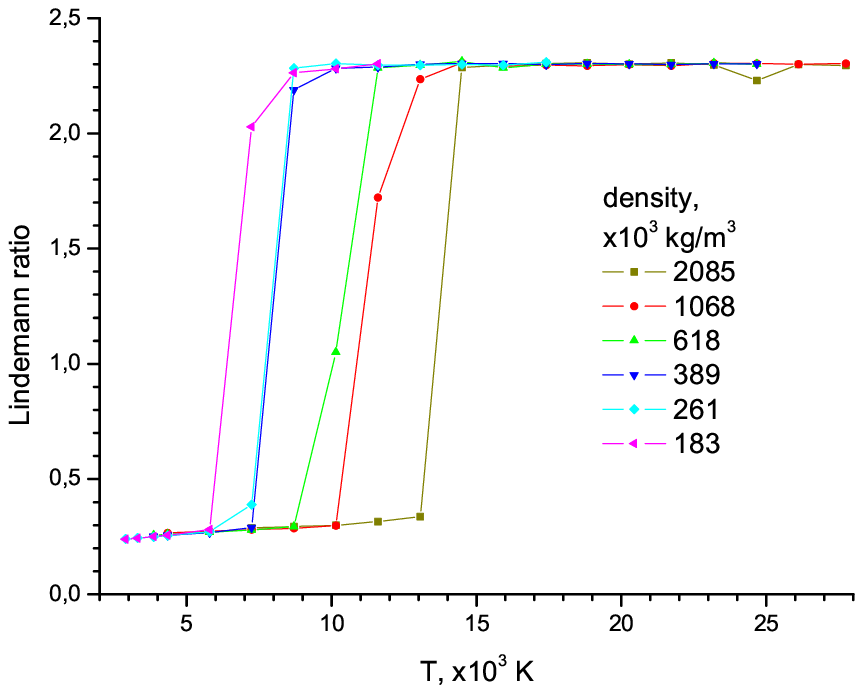}
\end{center}
\caption{Lindemann ratio at different densities}
\label{Lin_all}
\end{figure}

The Lindemann ratio (\ref{Lind_R}) is a good measure of disorder
of the lattice, so it is extremely useful and obvious to detect
the phase transition, where the order totally vanishes. Figure
\ref{Lin_all} shows Lindemann ratio for all the range of explored
densities and temperatures. The phase transition is clearly seen
here. Note that while the plateau in solid phase (left bottom)
gives some physical information, the plateau in liquid phase
(right top) is due to finite volume effects and it is determined
only by the volume.

\begin{figure}[h!]
\begin{center}
\epsfysize=100mm\epsfbox{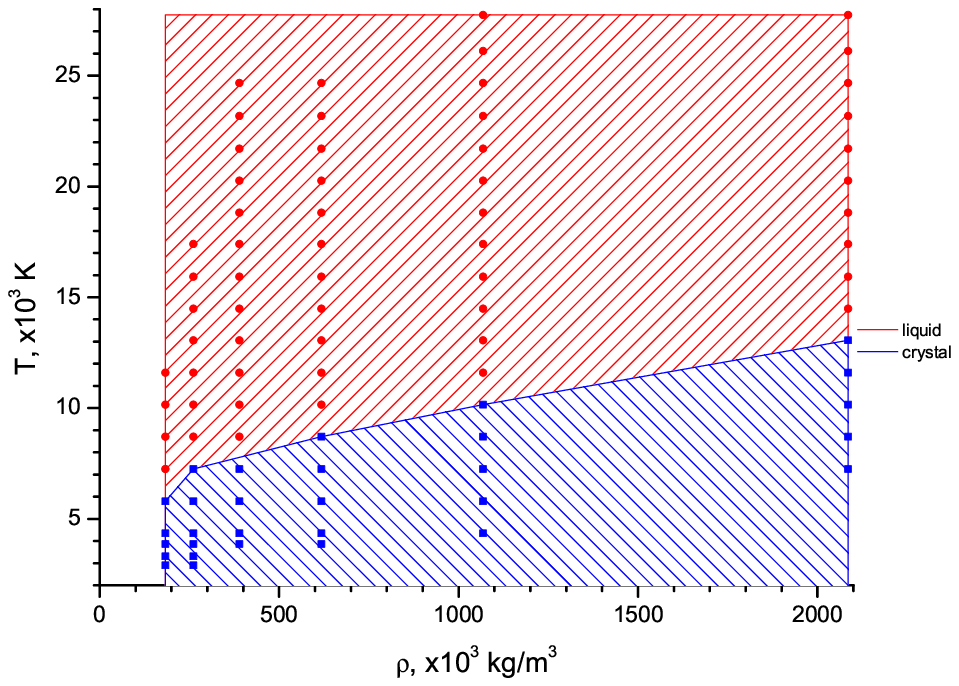}
\end{center}
\caption{Phase plane}
\label{phase_plane}
\end{figure}

The position of the phase transition is determined quite accurate,
so we can draw the phase plane for metal hydrogen. It it shown in
the Figure \ref{phase_plane}.

\begin{figure}[h!]
\begin{center}
\epsfysize=100mm\epsfbox{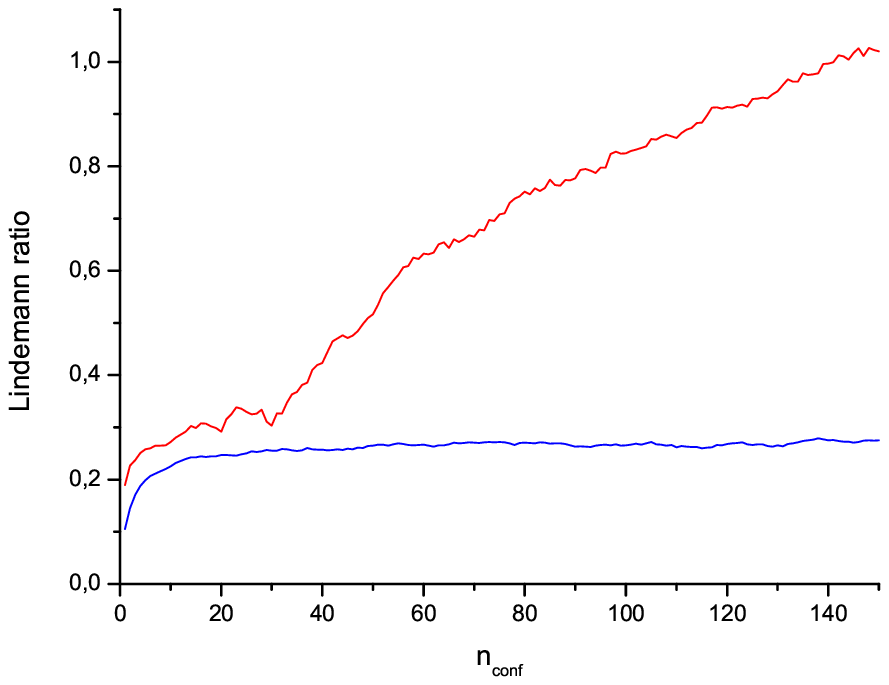}
\end{center}
\caption{Lindemann ratio thermalization at $\rho=2085\times10^3\
\mathrm{kg/m^3},\ T=13.1\ \mathrm{kK}$ (blue) and $T=14.5\
\mathrm{kK}$ (red)} \label{Lin_therm}
\end{figure}

Here we have to describe some important details. As we know, the
PIMC observables are averages over a set of thermalized path. To
get this set we start with any path and perform a Markov chain
procedure called thermalization. Sometime we start to get thermal
equilibrium paths, but we do not know how soon it will be. It is
well known that models of systems near a phase transition are
usually difficult to be thermalized over the transition. For
example we start our simulation with ideal "zero temperature"
crystal lattice (solid state). During the calculations it
thermalizes quite fast to some other solid state that seems
stable. It takes quite much calculation time to receive true
physical paths. An example is shown in Figure \ref{Lin_therm}.
This is the main obstacle to determine the position of the phase
transition more accurate. Moreover, it turns out that
thermalization from liquid to solid state takes so much time that
it hardly ever can be performed in moderate time. So, the position
of the phase transition is formally the upper limit. The lower
limit must formally be determined with a series of simulations
starting from "liquid" path. But it is not expected to differ much
from the upper limit that we received.

\section{Conclusions}

Path integral Monte Carlo technique was implemented to simulate atomic metal hydrogen from the first principles. Its thermodynamical properties were explored in a wide area of the phase plane. Numerical equations of state were obtained. The phase transition between liquid and solid crystal phases was detected and explored.

The principal thermodynamic parameters: temperature, density, pressure and energy were set or measured, but entropy was not. That will be the object of our following studies. The algorithms of obtaining entropy and adiabats are a little bit more complicated than for isoterms for example, because entropy can not be measured as a PIMC observable. So we have to solve differential equations derived from thermodynamics. Formally they give all information about the system since we know $E(\rho,T)$ and $P(\rho,T),$ but it is not trivial to get the numerical results. As it was mentioned, the lattice of calculation points in temperature and extremely in density must include close points in a large range that means much calculations. On the one hand we want to explore a wide range. On the other hand, the points must be close enough to allow the calculation of derivatives. We also plan to develop an alternative way of derivatives calculation based on constructing observables for them.

An other problem to be explored is to perform the thermalization from "liquid" to crystal solid state in order to determine the lower limit for the phase transition. We expect that it can be done much faster starting from two phase system.

\section{Acknowledgements}

The reported study was supported by the Supercomputing Center of
Lomonosov Moscow State University \cite{parallelru}.

This work was partially supported by The Ministry of education and
science of Russian Federation (grant No.8376).


\end{document}